# Porphyrin-functionalization of CsPbBrI$_2$/SiO$_2$ core-shell nanocrystals enhances the stability and efficiency in electroluminescent devices

*Jan Wahl, Manuel Engelmayer, Mukunda Mandal, Tassilo Naujoks, Philipp Haizmann, Andre Maier, Heiko Peisert, Denis Andrienko\*, Wolfgang Brütting\*, Marcus Scheele\**

J. Wahl, P. Haizmann, Dr. A. Maier, Prof. Dr. H. Peisert, Prof. Dr. M. Scheele
Institute of Theoretical and Physical Chemistry, Eberhard-Karls University of Tübingen, 72076 Tübingen, Germany
E-Mail: marcus.scheele@uni-tuebingen.de

Prof. Dr. M. Scheele
Center for Light-Matter Interaction, Sensors & Analytics LISA$^+$, Eberhard-Karls University of Tübingen, 72076 Tübingen, Germany

M. Engelmayer, T. Naujoks, Prof. Dr. W. Brütting
Institute of Physics, University of Augsburg, 86135 Augsburg, Germany
E-Mail: bruetting@physik.uni-augsburg.de

Dr. M. Mandal, Dr. D. Andrienko
Max Planck Institute for Polymer Research, 55128 Mainz, Germany
E-Mail: denis.andrienko@mpip-mainz.mpg.de

**Keywords**

Perovskites, Ligand Exchange, Electroluminescence, LEDs, Porphyrins, Nanocrystals

Surface ligand exchange on all-inorganic perovskite nanocrystals of composition CsPbBrI$_2$ reveals improved optoelectronic properties due to strong interactions of the nanocrystal with mono-functionalized porphyrin derivatives. The interaction is verified experimentally with an array of spectroscopic measurements as well as computationally by exploiting density functional theory calculations. The enhanced current efficiency is attributed to a lowering of the charging energy by a factor of 2–3, which is determined by combining electronic and optical measurements on a selection of ligands. The coupled organic-inorganic nanostructures are successfully deployed in a light emitting device with higher current efficacy and improved charge carrier balance, magnifying the efficiency almost fivefold compared to the native ligand.

## 1. Introduction

Lead-halide perovskite-based light-emitting devices (LEDs), both all-inorganic (*e.g.* $CsPbX_3$; X=Cl, Br, I) and hybrid (*e.g.* $MAPbX_3$; MA = methylammonium) variants, recently made significant advances with regards to efficiency.[1,2] Perovskite nanocrystals (NCs) are particularly promising for realizing LEDs due to their unique properties. Specifically, the large surface area of NCs offers numerous possibilities for functionalization with a variety of surface ligands. Therefore, a major challenge towards advancing nanocrystal-based LEDs is the search for a suitable functionalization. A suitable ligand can modify the NCs, and *vice versa*, in a variety of ways, which can be exploited to prepare novel materials that fit specific needs. This has previously been shown on well-characterized nanoparticle systems and is providing the foundation of the present work, which focuses on the enhancement of the optoelectronic properties by exchanging the native ligand shell with porphyrin-based ligands.[3–6] In the case of perovskite nanocrystals, a ligand exchange is inherently difficult due to the instability of the particles against polar environments. However, it holds the potential of improving crucial parameters like conductivity, stability and device efficiency.[1] In practice, this is realized by introducing a ligand with a higher binding affinity to the surface, which can lead to more stable NCs.[7] Additionally, the electronic coupling between NC and ligand can be adjusted and subsequently used to modify the optoelectronic properties at the interface.[3,5,8] Following this rationale, we introduce a series of novel surface ligands, *i.e.* the semiconducting 5-monocarboxyphenyl-10,15,20-triphenylporphyrin and its metalated analogs, in an attempt to improve both stability and current efficiency. Porphyrins have already shown improved efficiency in perovskite-based light harvesting devices as an intermediate layer, which makes them a suitable candidate for functionalized hybrid materials.[9,10] Here, we investigate how this surface ligand affects key parameters of $CsPbBrI_2$ NC-based LEDs, namely the current efficacy and turn-on voltage. To this end, we demonstrate how exchange with this porphyrin surface ligand becomes possible without deteriorating the perovskite nanocrystal. We show the effects of the porphyrin on the absorption, fluorescence and electrical transport in macroscopic thin films of the exchanged NCs and find a significant suppression of parasitic ionic transport, an increased crystal phase stability, reduced fluorescence lifetime as well as bright and narrow electroluminescence at 650 nm. In LEDs, these $CsPbBrI_2$ NC/porphyrin hybrid materials exhibit improved current efficacies, which we attribute to a reduction of exciton quenching and an improved charge carrier balance.

## 2. Results and Discussion

### 2.1. Structure and core-shell synthesis

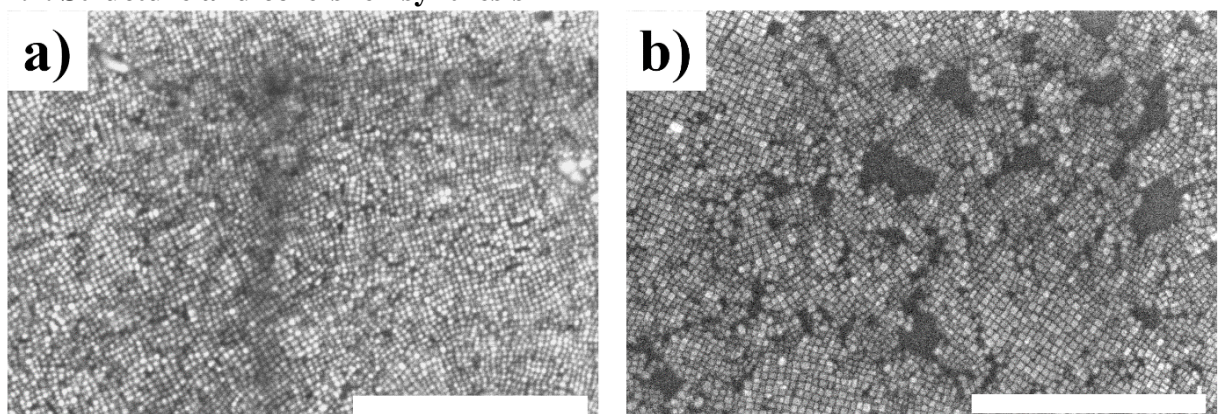

**Figure 1.** SEM micrographs of $CsPbBrI_2$ nanocrystals showing the preservation of the structural integrity during the preparation and ligand exchange processes. a) Untreated NCs without inorganic shell, and b) after the ligand exchange with mZnTPP. Scale bars correspond to 500 nm.

We find that exchange with ligands bearing polar functional groups on perovskite NCs is inherently difficult. To overcome this, we synthesized a silica shell ($SiO_x$) around the perovskite core to enhance its stability (Details in **Figure S3**).[12] To monitor the particle and superstructure morphology during the preparation process, SEM and AFM measurements were taken before and after the shell growth, which displayed homogeneous, well-defined thin films with a thickness of ~15–19 nm, corresponding to approximately one monolayer of NCs. By comparing the micrographs before and after the shell growth, we find an increase of the particle diameter from $d_0 = 8.9 \pm 1.2$ nm to $d_1 = 10.9 \pm 1.4$ nm. This suggests that a silica shell with a thickness of approximately 1 nm was synthesized (**Figure S1a–S1b**). Further micrographs, size distributions and spectroscopic analyses are given in the SI (**Figure S1 and S2**). The expected improvement in stability is especially beneficial for the functionalization with a polar ligand, like metal-(5-monocarboxyphenyl-10,15,20-triphenylporphyrin) (mMTPP, see **Figure 2b**). The semiconducting character and interaction as an intermediate layer in perovskite devices predestine this porphyrin as a promising surface ligand.[9,10] After functionalization with the organic semiconductor, no apparent changes in morphology could be observed as the particles kept their cubic shape and size (**Figure 1b**), indicating that the ligand preserves the structural integrity. The nearest neighbor distance did not change during the $SiO_x$-shell synthesis nor during the ligand exchange.

## 2.2. Ligand exchange

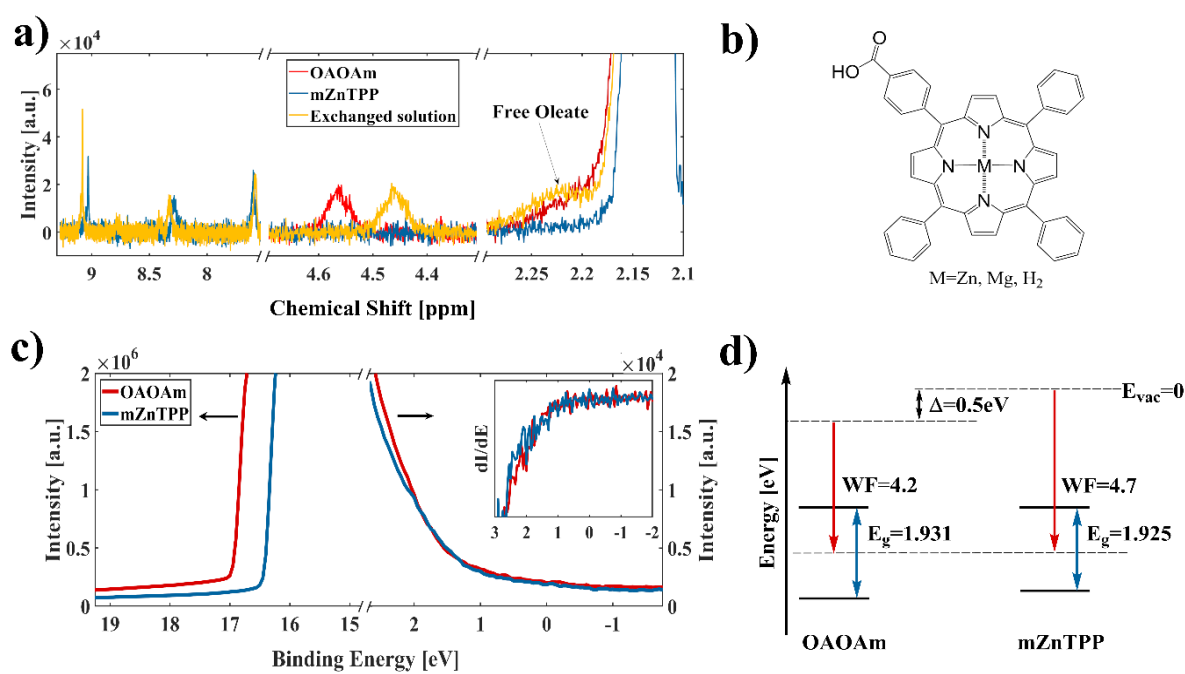

**Figure 1.** a) NMR analysis of the ligand exchange with mZnTPP. Shown are the regions for the characteristic porphyrin (7.5–9 ppm), amide proton (4.4–4.6 ppm) and the caesium oleate (2.2–2.25 ppm) peaks. b) Structure of the metal-(5-monocarboxyphenyl-10,15,20-triphenylporphyrin) derivatives described throughout this work. c) UPS of thin films before (blue) and after (red) ligand exchange. The inset displays the derivative. d) Corresponding

electronic structure of the films, shown are the work functions (red) and band gap energies (blue). The values are referenced against the instrument Fermi level. Overview spectra of the NMR and UPS measurements can be found in **Figure S4 and S5**.

The solution-based ligand exchange with different mMTPP derivatives (**Figure 2b**) was monitored by NMR (**Figure 2a**). Conveniently, a peculiarity of the iodide containing perovskites is the catalysis of amide formation,[13] which we use here to monitor the exchange. The characteristic amide peak at ~4.6 ppm experienced a clearly visible upfield shift, which is commonly attributed to a detachment of the native ligand.[14] Additionally, a downfield shift for the proton peaks in the characteristic porphyrin region (7.5–9 ppm) suggests successful binding of the desired ligand.

Since mMTPP is significantly more polar than oleic acid (OA) and oleylamine (OAm), the surface dipole moment is expected to change upon introducing the novel ligand, hence shifting the absolute energy levels depending on the strength and orientation of said dipole moment.[5] Since mMTPP has a pronounced π-system and an asymmetric functionalization, it provides an intrinsic dipole moment. This dipole is directed contrary to the surface dipole moment of the NCs and is thus able to partially compensate it. We find a shift of the Fermi level by 0.5 eV to lower energies in the case of mZnTPP (**Figure 2c–d**). In both cases, the Fermi level remained in the middle of the respective band gap, indicated by the same onset of the valence band maximum (**inset of Figure 2c**).

## 2.3. Optical properties

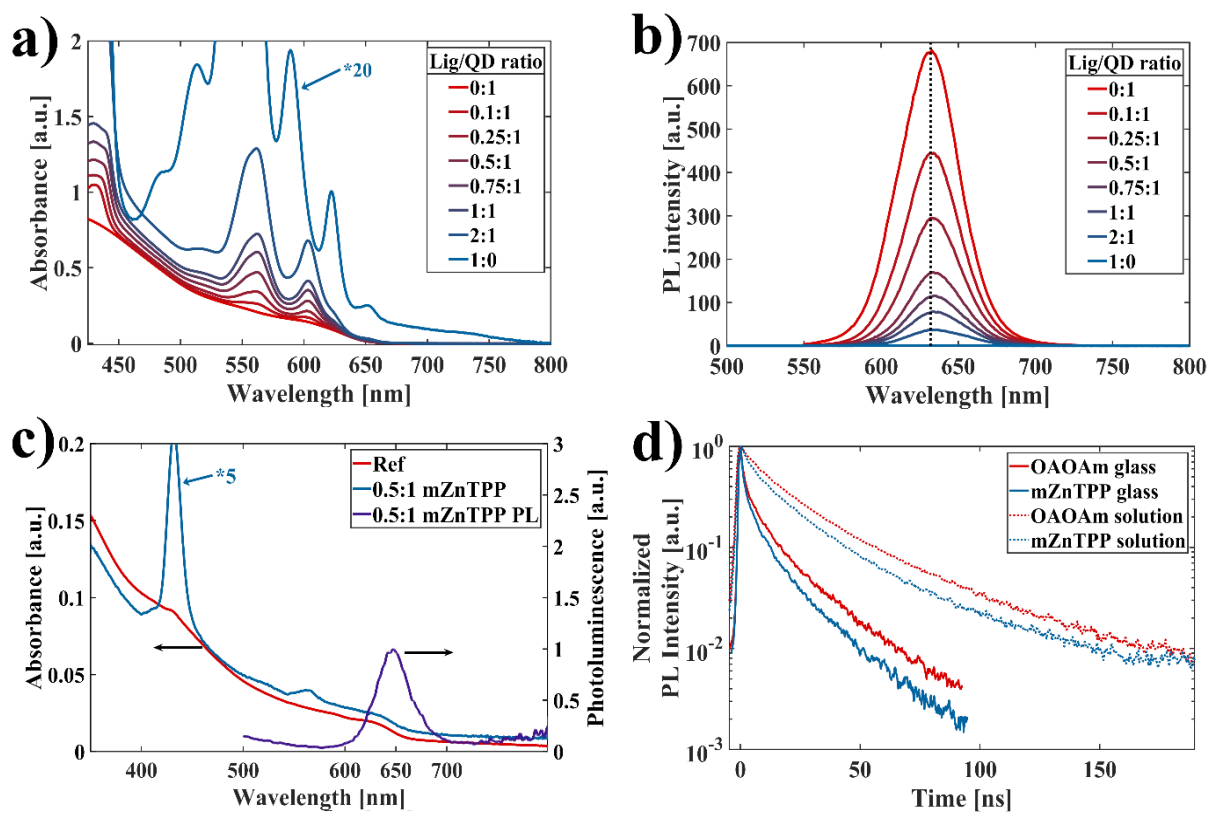

**Figure 3.** A) UV-vis and b) PL in toluene with different stoichiometric amounts of mZnTPP added. C) Solid state absorption and photoluminescence. D) Exciton lifetime measurement of native and exchanged samples in both solution and on a glass substrate.

The native $CsPbBrI_2$ sample exhibits the typical spectroscopic features for this composition and a quantum yield of ~60%. The absorption spectra (**Figure 3a**) confirm the preservation of the first excitonic transition from the valence state ($1S_h$) to the conduction state ($1S_e$) of the NCs as the onset of the absorption is independent of the amount of ligand added. Nevertheless, increasing the ligand content leads to more pronounced characteristic absorption peaks, namely the Q-bands of the porphyrin. Strikingly, these bands are shifted to lower energies when compared to an isolated mZnTPP sample. This can be attributed to a pronounced interaction between the particle and ligand since porphyrins are very sensitive towards electronic changes in their periphery.

By comparison with the photoluminescence (PL) spectrum of pure mZnTPP (spectra shown in SI, **Figure S6**), we verify that the emission of the mixed solution originates purely from the perovskite NCs. Upon addition of mZnTPP, we observe a redshift in the emission by 6 meV and a decrease in quantum yield (QY) as indicated by the reduced emission intensity (**Figure 3b**). Moreover, fluorescence lifetime measurements display 30% shorter lifetimes after ligand exchange (**Table 1**), indicating an enhanced interaction between particle and ligand. The redshift in the emission spectra can be explained by a spatial extension of the exciton wavefunction onto the ligand, a changed dielectric environment or a combination of both. Likewise, the shorter fluorescence lifetime can be rationalized as follows: by extending the excitonic wavefunction onto the ligand, the quantum confinement is lowered which results in decreased exciton binding energies.[15,16] The lowering destabilizes the exciton, yielding decreased PL lifetimes. This is consistent with the observed behavior in the solid state where the nearest neighbor interaction enhances the said effect, resulting in a bathochromic shift from 635 nm to 650 nm (**Figure 3c**) and a further reduction of the lifetime by a factor of 2–3 (**Figure 3d and Table 1**). Additionally, a separation of excitons can occur in the system, as the conduction state ($1S_e$) is shifted onto the ligand in the mZnTPP functionalized NCs (**see Figure 5d**). Long-lived excitons can thus be split and are able to recombine non-radiatively on the ligand. Especially porphyrins are known for providing a variety of radiationless transitions.[17–19] This scenario would also invoke a decrease in lifetime and QY.

**Table 1.** Measured photoluminescence lifetimes ($\tau_i$) of $CsPbBrI_2$ samples on glass and in solution. The values were obtained by fitting the PL decay with a triexponential function. The goodness of the fit is given as the R-squared factor ($R^2$).

|  | Solution | | | | Solid state | | | |
| --- | --- | --- | --- | --- | --- | --- | --- | --- |
| Ligand | Lifetime $\tau_i$ [ns] | | | $R^2$ | Lifetime $\tau_i$ [ns] | | | $R^2$ |
|  | $\tau_1$ | $\tau_2$ | $\tau_3$ |  | $\tau_1$ | $\tau_2$ | $\tau_3$ |  |
| OAOAm | 2.89 | 13.82 | 39.27 | 0.99985 | 1.09 | 6.59 | 22.46 | 0.99971 |
| mZNTPP | 2.25 | 12.93 | 34.06 | 0.99983 | 0.84 | 4.80 | 17.34 | 0.99957 |

## 2.4. Electronic properties

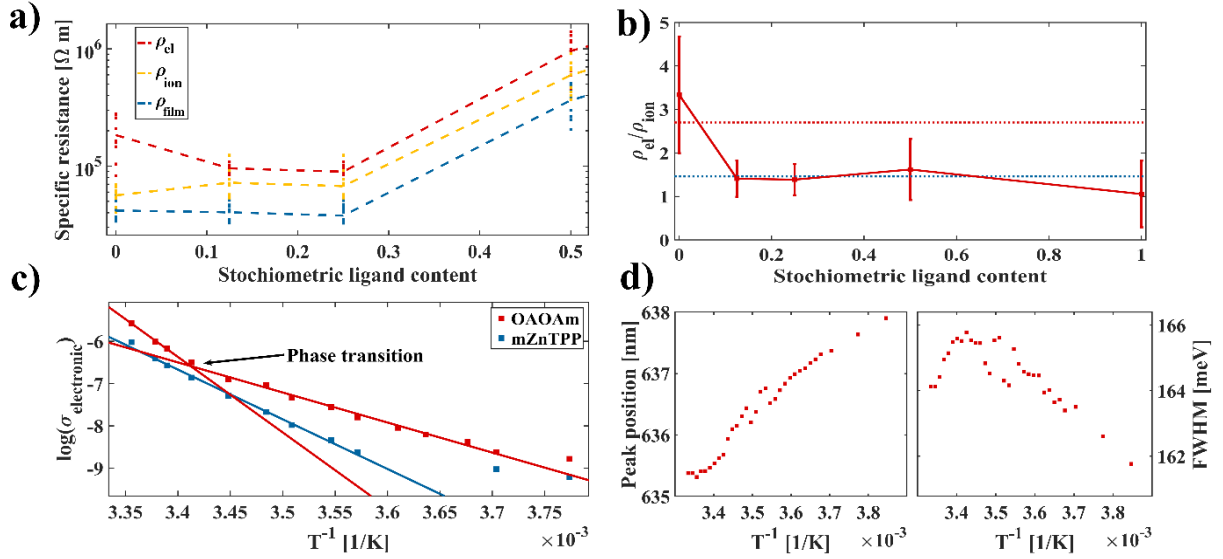

**Figure 4.** a) Specific resistances of thin films with different ligand compositions. The electronic (red) and ionic (orange) contributions to the overall ohmic resistance (blue) are shown. b) Ratio of electronic and ionic resistance at different stoichiometric values. Dotted lines represent the ratio of electronic and ionic activation energies obtained from temperature dependent conductivity measurements. c) Temperature dependence of the electronic contribution with the corresponding fitting functions. The ionic contribution is given in **Figure S7** d) Temperature dependent PL measurements. Position (left) and FWHM (right) of the PL peak in dependence of temperature.

Since perovskites are both electronic and ionic conductors, both transport pathways need to be considered. This can be effectively described by a parallel connection of an electronic ($\rho_{el}$) and ionic resistance ($\rho_{ion}$).[20,21]

$$\frac{1}{\rho_{film}} = \frac{1}{\rho_{el}} + \frac{1}{\rho_{ion}} \tag{1}$$

To derive $\rho_{el}$ from **Equation (1)**, the ionic part has to be eliminated. Since the ionic contribution behaves like a capacitor, it can be saturated by a potentiostatic measurement, *i.e.* if a constant voltage over a prolonged period of time is applied. By combining a standard current-voltage curve to obtain the film resistance ($\rho_{film}$) and the potentiostatic measurement for $\rho_{el}$, the specific resistances with varying ligand content were analyzed (**Figure 4a**). The resistance of the system increases dramatically if 0.5 or more equivalents of mZnTPP are added. The most probable reason is a saturation of the NC surface and the allocation of free, excess ligand between the particles. At lower stoichiometric additions, the ratio $\rho_{el}/\rho_{ion}$ decreases, while the overall film resistance stays constant (**Figure 4b**). To validate these findings and further explore the conduction mechanism, temperature dependent conductivity measurements were carried out. The results were fitted using an Arrhenius type nearest neighbor hopping (NNH) model:

$$\sigma(T) = \sigma_0 exp\left\{-\frac{E_A}{k_B T}\right\}, \tag{2}$$

with the conductivity $\sigma_0$, Boltzmann's constant $k_B$, temperature $T$ and the hopping activation energy $E_A$. Upon mZnTPP-functionalization, the activation energy in the exchanged system was reduced by over 33% for the electron hopping, whereas the energy for the ion transport did not change significantly (**Figure 4c, Table 2 and Figure S7**). Which further supports our claim of improved electrical properties of the mZnTPP ligand system.

**Table 2.** Results of the Arrhenius type NNH fit of the temperature dependent conductivities. The electronic activation energy ($E_A^{el}$) and ionic activation energy ($E_A^{ion}$) were measured. In the case of the native OAOAm ligand shell the higher temperature fit was used. The ratio of the energies was compared to the measured specific resistance ratio (SRR) as shown in **Figure 4b.**

|  | OAOAm | mZnTPP |
|---|---|---|
| $E_A^{el}$ [eV] | 3.56 | 2.34 |
| $E_A^{ion}$ [eV] | 1.38 | 1.41 |
| $E_A^{el}/E_A^{ion}$ | 2.57 | 1.65 |
| $\rho_{el}/\rho_{ion}$ | 3.2 | 1.5 |

We note that the temperature-dependent conductivity of the native system cannot be described with a single linear fit, which we tentatively attribute to a phase transition at around 290 K. To verify this assumption, the PL was additionally monitored over the same temperature range (**Figure 4d**). Two observations were evident, the emission peak shifted to lower energies upon cooling the sample and the full width at half maximum (FWHM) exhibited a maximum at 290K.

The shift arises due to the thermal compression of the lattice and an increased interaction between lead and halides, resulting in a broadening of states at the valence band edge which reduces the size of the band gap, this is consistent with previous reports on perovskites.[22,23] Further, as the FWHM of the PL signal is phonon-dependent and typically increases with temperature $FWHM \propto exp\{E_{LO}/(k_B T)\}^{-1}$, a necessity of the decrease in broadening with higher temperature is a phase transition. Tang *et al.* have investigated the electron-phonon coupling of different perovskite phases and their temperature dependent behavior.[23] They found the energy of the longitudinal optical phonon in cubic nanoparticles to be two times larger than in the corresponding orthorhombic counterparts, therefore, a transition from predominantly cubic (T > 290 K) to orthorhombic (T < 290K) could explain the decrease of FWHM as observed in **Figure 4**.[24,25]

## 2.5. Charging energy

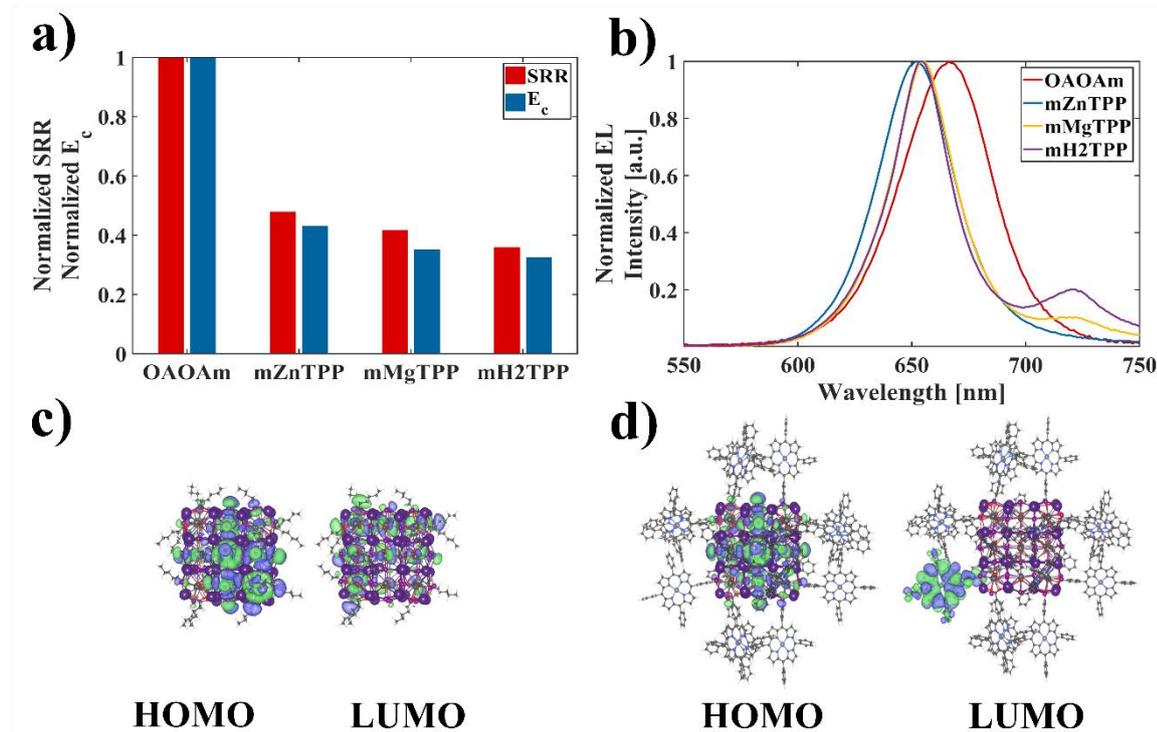

**Figure 5.** a) Normalized charging energy ($E_C$) and measured specific resistance ratios (SRR) for OAOAm and mMTPP (M = Zn, Mg, H$_2$). b) Electroluminescence spectra of devices built with differently functionalized NCs. Note that samples prepared with mMgTPP and mH2TPP

exhibit parasitic emission at ~725 nm. Frontier molecular orbitals of CsPbBrI$_2$ NC decorated with c) hexanoic acid as a mimic for oleic acid and d) mZnTPP ligands.

To clarify the origin of the altered optoelectronic properties, the dependence of the conductivity on the charging energy $E_C$ was considered, which describes the energy required to introduce charges in a nanocrystal. Following the modified Laikhtman-Wolf model $E_C$ is defined as: [26]

$$E_C = \frac{e^2}{2(C_S + n\, C_m)}, \qquad (3)$$

with the elementary charge $e$, self-capacitance $C_s$, number of nearest neighbors $n$ and the mutual capacitance $C_m$. A more detailed derivation of the components can be found in the SI (**section S9**).

**Table 3.** The calculated molecular polarizabilities of the different ligands in atomic units [au], OAOAm refers to the native system, mZnTPP, mMgTPP, mH2TPP refer to the zinc, magnesium and metal free derivative, respectively. The calculated dielectric constants as well as the charging energies **(Equation 3)** are shown as well.

|        | Polarizability [au] | Dielectric constant | Charging energy [meV] |
|--------|---------------------|---------------------|------------------------|
| OAOAm  | ~240                | 2.5                 | 8.62                   |
| mZnTPP | 808                 | 4.4                 | 4.38                   |
| mMgTPP | 809                 | 4.9                 | 3.72                   |
| mH2TPP | 797                 | 5.1                 | 3.49                   |

To evaluate the main influence of the charging energy on the charge carrier dynamics, the NCs were functionalized with mMTPP derivatives. Upon comparing the measured, normalized specific resistance ratios (SRR) with the charging energy $E_C$ of the different systems, it becomes clear that the improvement of the electronic properties follows the same trend as $E_C$ (**Figure 5a**), suggesting that the charging energy has the main influence. Quantitatively, the charging energy of the exchanged samples, displayed in **Figure 5a**, is lowered by a factor of ~2.4–3 compared to the native NCs. We note that although mMgTPP and mH2TPP showed beneficial electrical properties compared to mZnTPP, such as a better SRR and lower $E_C$, they were found unsuitable for light emitting devices due to parasitic emission at longer wavelength (**Figure 5b**).

To probe the effect of the ligand used for surface functionalization on the electronic structure of the NC/ligand hybrid system, density functional calculations were carried out (see SI, **section S10** for computational details). As shown briefly in **Figure 5c–d**, and in the supporting information (**Figure S9–S12**) in more detail, a significant change in the nature of the frontier orbital can be observed as a function of ligand. For the aliphatic hexanoic acid ligand we find that, both occupied and empty molecular orbitals (MOs) reside over the inorganic core (**Figure 5c**). The introduction of the porphyrin ligand and its metalated derivatives, however, changes this significantly, and we see that the low-lying vacant orbitals are exclusively over the porphyrin moiety (**Figure 5d**). This shift of the $1S_e$ state onto the ligand and hence reduced exciton confinement can lead to the aforementioned reduction in PL lifetime, and a decrease in surface dipole moment (compare **Figure 2**). Considering the porphyrin-functionalized NCs on themselves, we see that in the case of the mZnTPP system, the occupied orbitals H, H–1, and H–2 (H ≡ highest occupied MO or HOMO) reside almost entirely over the NC core (**Figure S9**). However, the scenario changes once we switch to the other two porphyrin-based systems. Accordingly, even for the occupied MOs, we find significant density over the ligand for both the mMgTPP (see the H-1 and H-2 orbitals; **Figure S10**) and the mH2TPP system (refer to the H, H–1 and H–2 orbitals; **Figure S11**). We hypothesize that this partial extension of the occupied orbital over the ligand can contribute towards the slightly favorable optoelectronic properties of mMgTPP and mH2TPP compared to those of mZnTPP. However, we hold these high-lying, ligand-based occupied orbitals in mMgTPP and mH2TPP responsible for the parasitic emission observed near 725 nm, more so since the low-lying unoccupied orbitals, i.e. the $1S_h$ state, is always located over the ligand in the porphyrin-exchanged systems.

## 2.6. Optoelectronic devices

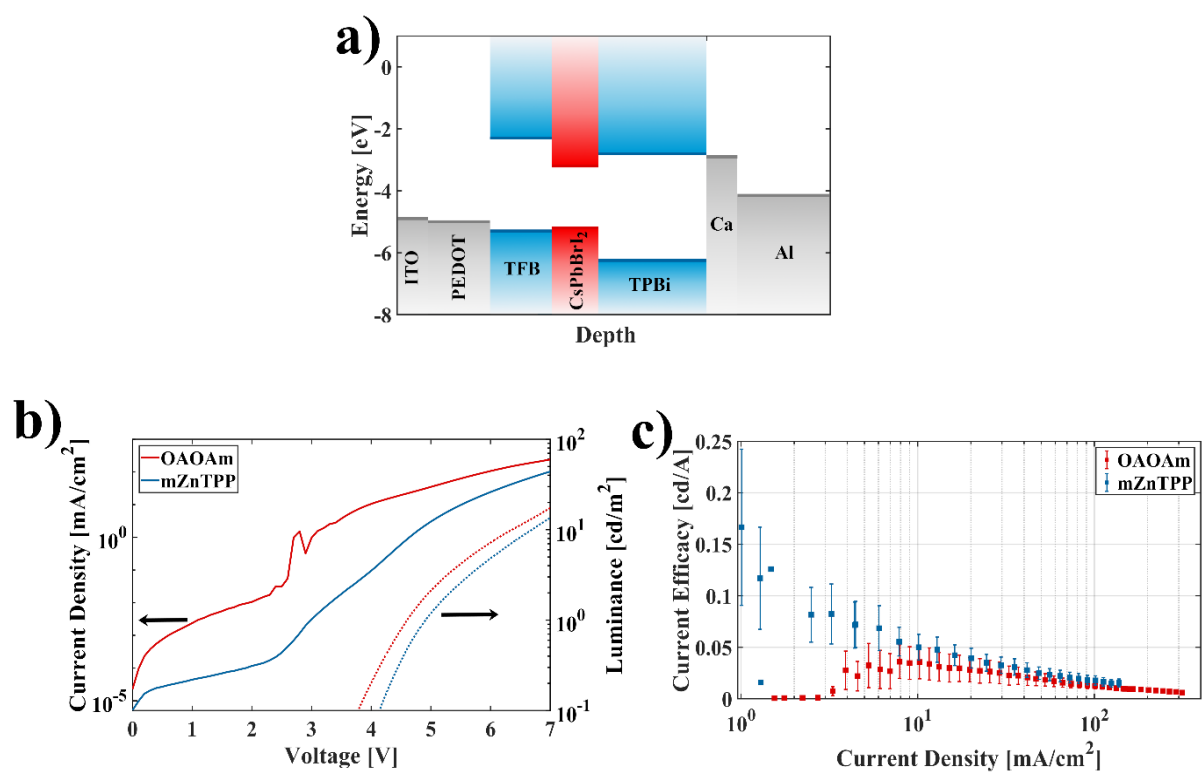

**Figure 6.** a) Typical stack design used for the characterization of CsPbBrI$_2$-based light emitting devices. b) Typical current-voltage-luminance (jVL) characterization curve, with the current density and luminance shown as a solid and dashed line, respectively. c) Measured current efficacies of native and exchanged perovskites in electroluminescent light emitting devices. The efficiency curves were obtained by averaging over five different samples, each consisting of four pixels.

The perovskite NCs with and without ligand exchange were deposited as the emitter material in the device stacks, shown in **Figure 6a**, with a homogeneous perovskite NC film as an emitting layer (see AFM image in **Figure S13**). We obtained electroluminescent devices with a turn-on voltage of the current of ~2.5 V (**Figure 6b**) and a turn-on voltage of the luminance in the range of 3–4.5 V. From this comparison, we find that the ligand exchanged NCs exhibit almost the same luminance, but at lower current densities, which results in an improved current efficacy (**Figure 6c**). Overall, the peak current efficacy could be enhanced by a factor of 4.65 and the necessary current density to reach the peak efficacy was also lowered by approximately one order of magnitude. This could indicate that the ligand exchange leads to reduced quenching of excitons, *i.e.* the injected charge carriers are not lost during injection. Additionally, we find that the maximum current efficacy for the systems was located either at the onset or at higher current densities for mZnTPP and OAOAm, respectively. A possible explanation is an improved charge carrier balance in the exchanged system. We attribute both phenomena to the lower charging energy in the porphyrin-functionalized system.

## 3. Conclusion

To summarize, we successfully performed ligand exchange with mMTPP derivatives on all-inorganic core-shell $CsPbBrI_2/SiO_x$ nanocrystals with a core size of 9 nm while maintaining the structural integrity of the particles. We found the optoelectronic properties of the system to be highly controllable with the nature and amount of ligand, suggesting significant electronic interaction between nanoparticle and ligand. This is also evidenced by a decrease of the first excitonic transition energy and photoluminescence lifetimes, where a dislocation of the $1S_e$ state onto the ligand could be identified as the main driving force. Additionally, the composition of the measured current, namely the electronic and ionic contribution, was found to be dependent on the dielectric constant of the ligand. We hold a decrease of the charging energy and the manifestation of a compensating dipole moment at the NC interface responsible for the observed smaller electrical resistance. Temperature-dependent photoluminescence measurements indicated an increased phase stability of the ligand-exchanged perovskite nanocrystals. These improvements of the optoelectronic properties could be used to prepare light-emitting devices with enhanced current efficacy.

## 4. Experimental section

*Materials:* 1-Octadecene (ODE), technical grade, 90%, Sigma Aldrich; Oleic acid (OA), 97%, Acros Organics; Oleylamine (OAm), 80–90%, Acros Organics; Caesium carbonate ($Cs_2CO_3$), 99.99% (trace metal basis), Acros Organics; Lead(II)iodide ($PbI_2$), 99.999% (trace metal basis), Sigma Aldrich; Lead(II)bromide ($PbBr_2$), ≥98%, Sigma Aldrich; Tetramethyl orthosilicate (TMOS), 99%, Acros Organics; Toluene, HPLC grade, 99.8%; Toluene, 99.8%, extra dry, AcroSeal, Acros Organics; Ethyl acetate, ACS reagent, ≥99.5%, Sigma Aldrich; zinc-(5-monocarboxyphenyl-10,15,20-triphenylporphyrin) (mZnTPP), TriPorTech; magnesium-(5-monocarboxyphenyl-10,15,20-triphenylporphyrin), 98%min, PorphyChem; (5-monocarboxyphenyl-10,15,20-triphenylporphyrin), 98%min, PorphyChem; poly(3,4-ethylenedioxythiophene) polystyrene sulfonate (PEDOT:PSS) (AI4083 or CH8000), Clevios™, Heraeus Epurio; poly[(9,9-dioctylfluorenyl-2,7-diyl)-co-(4,4'-(N-(4-sec-butylphenyl)diphenylamine)] (TFB), $M_W$=10,000-30,000, Lumtec; 1,3,5-Tris(1-phenyl-1HBenzimidazol-2-yl)benzene) (TPBi), >99.5% (HPLC), Lumtec

*Synthesis:* The $CsPbBrI_2$ nanocrystals in this work are prepared by following the literature on arrested metathesis reported by Protesescu *et al.* with slight modifications.[11] In a typical synthesis, two precursors were prepared in separate flasks, one containing 203.5 mg (0.624 mmol) $Cs_2CO_3$ and 0.7 ml oleic acid (OA) in 10 ml octadecene (ODE). The second flask contained 116 mg (0.252 mmol) $PbI_2$ and 46 mg (0.125 mmol) $PbBr_2$ in 5 ml ODE, both flasks were dried under vacuum at 120 °C for 2 h. Immediately after the addition of OA to the first flask, gas evolution was observed due to the formation of carbonic acid which decomposes to water and carbon dioxide, ultimately indicating the formation of caesium oleate. After completion of the drying process, the flasks were set under nitrogen.

The caesium oleate flask was then heated to 150 °C to ensure completion of the caesium oleate formation, visually observable by a yellow-brownish color. Before injection, said solution was cooled to 120 °C. To dissolve the lead salts, 1 ml OA and 1 ml OAm was added to the flask, the solution was heated to 160 °C after the salts were dissolved. Subsequently, 0.8 ml of caesium oleate precursor were quickly injected into the lead halide solution under vigorous stirring. After allowing the reaction to proceed for 5 s, the reaction was quenched by placing it in an ice-bath.

The purification was carried out *via* centrifugation at 4000 rpm for 12 min, the pellet was afterwards redispersed in dry toluene under nitrogen atmosphere.

The general procedure for the silica shell was based upon the procedure published for CsPbBr$_3$ particles.[12] The thickness of the as synthesized shell could be tuned by using different amounts of tetramethoxyorthosilicate (TMOS), commonly a ratio of 5:1 (TMOS:NC) was used. A 1:1 mixture of wet and dry toluene was used as a solvent, since water is the initiator for the polymerization. After 72 h reaction time, the solution was again centrifuged at 2000 rpm for 2 min to remove unwanted larger particles and agglomerates. The supernatant was subsequently transferred to a fresh vial to obtain a stable colloidal nanoparticle dispersion. If necessary, the nanoparticle dispersion was afterwards filtered with a 200 nm syringe filter.

*Ligand exchange:* The ligand exchange was carried out in solution by utilizing a defined solution of metal-(5-monocarboxyphenyl-10,15,20-triphenylporphyrin) (mMTPP) in toluene and adding specific volumina to the NC dispersion. Upon addition, an immediate color change was observed.

The solutions, exchanged as well as native, were then used as prepared or washed with ethyl acetate (EtAc), depending on the desired analysis method. The washing step was carried out by adding 2 volumetric equivalents of EtAc to the NC dispersion followed by centrifugation. The resulting precipitate was redispersed in toluene.

*Spin-coating and vacuum drying:* Most characterization techniques presented in this work are carried out in thin films of the as prepared NCs. The method of choice was spin–coating under inert atmosphere. Usually, 100 µl of a ~2 mM solution (particle concentration ~0.7 µM) was casted onto a substrate (silica, gold or device stack) and left at rest for 30 s. The substrate was spun at 600 rpm for 30 s with a ramp of 3 s. Since low spinning speeds were used, the spin-coater was repeatedly started 2–3 times after the initial 30 s to remove any residual solvent which remained at the edges. Subsequently, the substrate was placed in vacuum for 15 min for drying.

*Conductivity measurements:* Conductivity measurements were carried out on 15×15mm$^2$ silicon field-effect transistor (FET) substrates with gold contacts of different channel lengths 2.5 µm, 5 µm, 10 µm and 20 µm obtained from the Fraunhofer IPMS, Dresden. Prior to the film preparation, the substrates were rinsed with acetone and subsequently cleaned with acetone, deionized water and ethanol in an ultrasonic bath for 5min each. Following the film preparation, the substrates were mounted on a custom build sample holder under inert atmosphere which was connected to a Keithley 2634B SYSTEM SourceMeter® with $10^{-9}$ A (nA) accuracy. Temperature dependent conductivity measurements were carried out inside a Lake Shore Cryotronics CRX-6.5K Probe Station attached to a Keithley 2636B SYSTEM Source Meter® with $10^{-12}$ A (pA) accuracy. In a typical film characterization, custom measurement scripts were used.

*Scanning electron microscopy (SEM):* SEM was carried out with a HITACHI SU8030 electron microscope operating with an acceleration voltage of 30 kV.

*Ultraviolet electron spectroscopy (UPS):* UPS measurements were performed using a multi-chamber UHV system with a base pressure of $2 \cdot 10^{-10}$ mbar, equipped an Ultraviolet source (UVS 300 SPECS) and a Phoibos 150 hemispherical photoelectron analyzer with DLD detector. Measurements were carried out with Helium I radiation (21.22 eV). The films for UPS were prepared on silicon/silicon oxide or gold substrates.

*Nuclear magnetic resonance (NMR) spectroscopy:* NMR measurements were performed on a Bruker Avance III HDX 400, with a frequency of 400 MHz. Spectra are analyzed with the Bruker TopSpin 4.0.2 software and plotted with MatLab R2020.

*UV-vis and photoluminescence spectroscopy (in solution and solid state):* Optical measurements were performed on a UV-vis-NIR spectrometer (Agilent Technologies, Cary 5000) and a fluorescence spectrometer (PerkinElmer FL8500). Both systems contain interchangeable sample holders for measurements in solution and solid state. The fluorescence spectrometer additionally provides the possibility of absolute quantum yield measurements in an integrating sphere.

*Fluorescence lifetime measurements:* The samples were excited *via* a picosecond pulsed laser system (EKSPLA PT400) and measured with an Acton Spectra-Pro 2300i spectrograph coupled to a Hamamatsu Streak Camera C5680 for time-resolved spectral information.

*Device characterization and current-voltage-luminance (jVL) curves:* The electroluminescent devices were prepared on prestructured indium tin oxide (ITO) which was cleaned with acetone and isopropanol in an ultrasonic bath. Subsequently, a layer of poly(3,4-ethylenedioxythiophene) polystyrene sulfonate (PEDOT:PSS) was spin-coated. On top, a layer of poly[(9,9-dioctylfluorenyl-2,7-diyl)-co-(4,4'-(N-(4-sec-butylphenyl)diphenylamine)] (TFB) was applied, a solution of 5 mg/ml in chlorobenzene was spin-coated at 3000 rpm and dried at 175 °C for 30 min. The perovskite layer was prepared as described before. Above the perovskite emitters, a layer of 1,3,5-Tris(1-phenyl-1HBenzimidazol-2-yl)benzene) (TPBi) was deposited. As a top electrode, calcium and aluminium were thermally evaporated onto the sample. The substrates are divided to 4 subsections, each with an active area of 2×2mm$^2$. The devices were measured in a custom-made sample box attached to a calibrated photodiode for luminance, a two-channel Keithley 2602B source-measure unit for powering the LED and probing the electrical as well as optical output, and a JETI specbos 1211 for spectroscopic measurements. Data acquisition was carried out with a custom written LabView program.

*Atomic force microscopy (AFM):* AFM was carried out with a Bruker MultiMode 8-HR and a Bruker Dimension Icon, the obtained images were analyzed using Gwyddion.

*DFT calculations:* All computations are performed using the CP2K 8.1 program suite[27] employing the PBE exchange correlation functional[28], MOLOPT[29] DZVP basis-set and GTH pseudopotentials[30] for core electrons. A dual basis of localized Gaussians and plane waves (GPW) with a 350 Ry plane-wave cutoff are used for all calculations. Grimme's DFT-D3 protocol[31] was used to account for van der Waals (VDW) interaction. SCF convergence criterion was set at $10^{-6}$ for all calculations.

Initial geometries of CsPbBrI$_2$ nanocrystals were obtained by cutting small cubes (~1.7 nm) from the bulk, exposing the CsX layer (X = Br, I) at the surface and maintaining overall charge neutrality of the particle.[32] All calculations invoked a periodic boundary condition. However, to avoid spurious interaction with its periodic image, nanoparticles were placed inside a large box of size 75×75×75 Å$^3$ ensuring large vacuum layer above the surface of the NC. All structures were then optimized in vacuum using the BFGS optimizer, setting a maximum force of 5 meV Å$^{-1}$ ($1.0\times10^{-04}$ hartree/bohr) as convergence criteria.

The polarizabilities of native ligands as well as the different metal-(5-monocarboxyphenyl-10,15,20-triphenylporphyrin) derivatives was carried out at the PW91/def2-TZVP level with the ORCA software.[34] The geometries were optimized using a SCF convergence criterion of $10^{-7}$.

**Supporting Information**
Supporting Information is available from the Wiley Online Library or from the author.
Cartesian coordinates of all computed structures can be accessed from the coordinate file (.xyz).

**Acknowledgements**
This work was supported by the DFG under grants SCHE1905/8-1, AN680/6-1, BR1728/21-1 (project no. 424708673) and SCHE1905/9-1. The authors would like to thank Dr. Kai Braun from the Institut für Physikalische und Theoretische Chemie, Universität Tübingen for his help with the measurements of the temperature-dependent photoluminescence spectra.

**Conflict of Interest**

The authors declare no conflict of interest.

# Supporting Information

# Porphyrin-functionalization of CsPbBrI$_2$/SiO$_2$ core-shell nanocrystals enhances the stability and efficiency in electroluminescent devices

*Jan Wahl, Manuel Engelmayer, Mukunda Mandal, Tassilo Naujoks, Philipp Haizmann, Andre Maier, Heiko Peisert, Denis Andrienko, Wolfgang Brütting, Marcus Scheele*

[1] *Institut für physikalische und theoretische Chemie, Universität Tübingen, Auf der Morgenstelle 18, D-72076 Tübingen, Germany*
[2] *Max Planck Institute for Polymer Research, Ackermannweg 10, D-55128 Mainz, Germany*
[3] *Institute of Physics, University of Augsburg, D-86135 Augsburg, Germany*
[4] *Center for Light-Matter Interaction, Sensors & Analytics LISA$^+$, Universität Tübingen, Auf der Morgenstelle 15, D-72076 Tübingen, Germany*

**S1. SEM and HRTEM micrographs of silica coated CsPbBrI$_2$ nanocrystals**

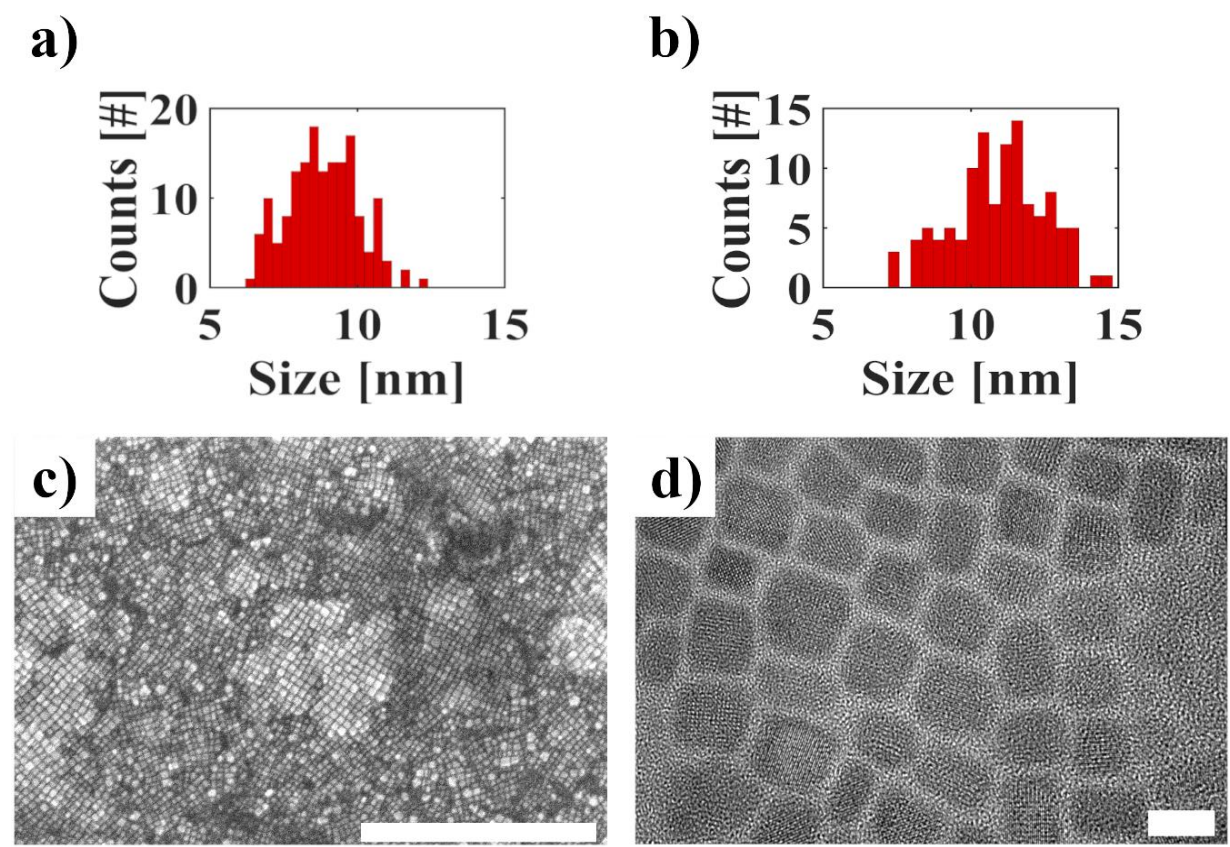

**Figure S2.** a) Size distribution determined from SEM micrographs of untreated CsPbBrI$_2$ nanocrystals. b) Size distribution after applying a 5:1 molar ratio of TMOS. c) SEM micrograph of silica coated NCs. d) HRTEM of silica coated NCs. The scale bars are 500 nm and 10 nm respectively.

## S2. AFM of spin-coated perovskite thin films with corresponding height determination

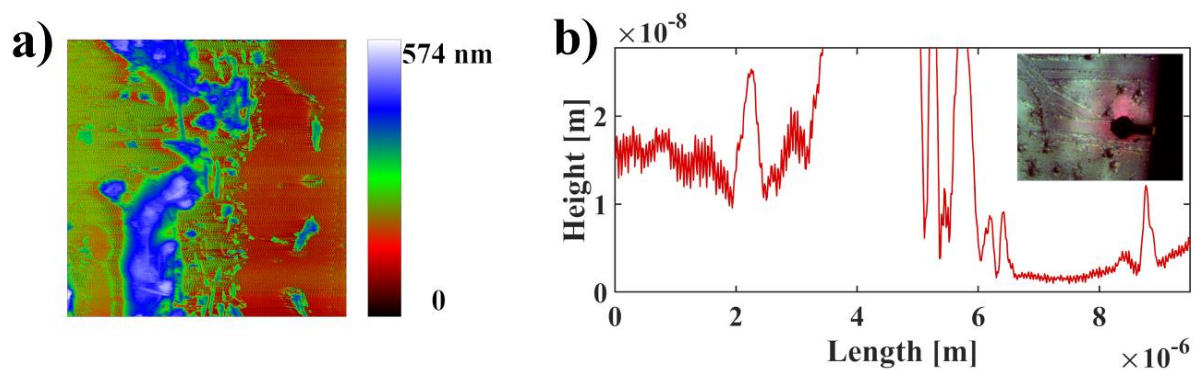

**Figure S3.** a) AFM image of a typical NC thin film, on the left the unharmed film is shown, whereas on the right a scratch was made to measure the absolute film height. The size of the micrograph is 10μmx10μm. b) Corresponding height profile of the NC thin film. The inset shows the different areas (scratch and unharmed) in a microscopic image. The large peak in between is due to dislocated material during the scratching process.

In order to determine the film thickness of a typical NC thin film, a film was prepared following the as described spin-coating method. Subsequently, a scratch was made with a syringe to remove the film in a certain area. The AFM tip was located onto the edges of the scratch and the measurement range was set to include the scratch as well as the unharmed thin film. Thereafter, the film was analyzed using Gwyddion and the film thickness was determined to be 15-19nm, which corresponds to approximately one monolayer.

## S3. Silica shell characterization and stability against polar solvents

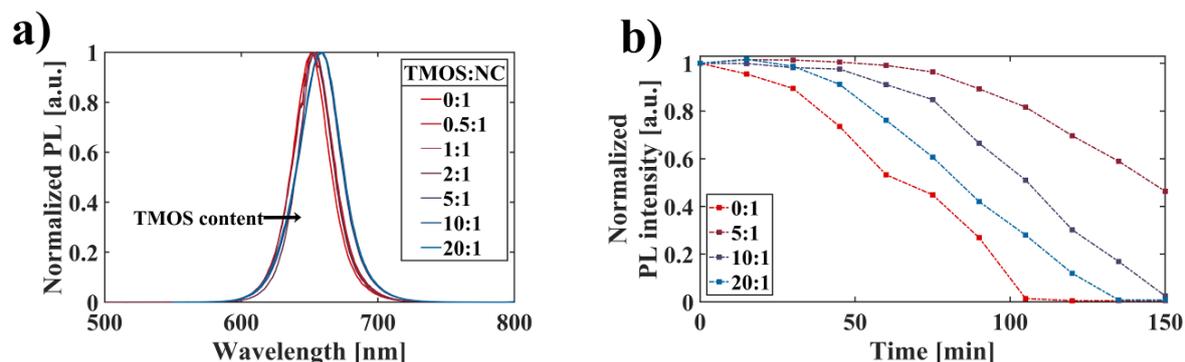

**Figure S4.** a) Normalized PL spectra of CsPbBrI$_2$ nanocrystals that were treated with different amounts of TMOS. The stochiometric ratio was ranging up to 20-fold. b) Stability of spectroscopic samples against water. In a typical experiment, 100 µl deionized water was added to a sample with spectroscopic concentration.

A hypochromic shift due to exciton leakage was observed when treating the CsPbBrI$_2$ samples with TMOS over the duration of 72 h. The extend of the shift is dependent on the amount of TMOS added and can be directly correlated to the shell thickness. The shift increases until the stochiometric ratio of 5:1 (TMOS:NC) is applied, adding additional TMOS does not shift the emission to lower energies.

The silica shell was subsequently probed for improved stability by adding 100µl deionized water to a spectroscopic sample with µM concentrations. It was found that the stability was maximized for a sample prepared with a 5:1 TMOS:NC ratio. Higher TMOS concentrations lead to a less stable silica shell, a possible explanation are incompletely reacted TMOS species that are activated upon water addition and hence form another highly reactive, polar species in solution which damage the particles.

## S4. Overview spectra of NMR and UPS

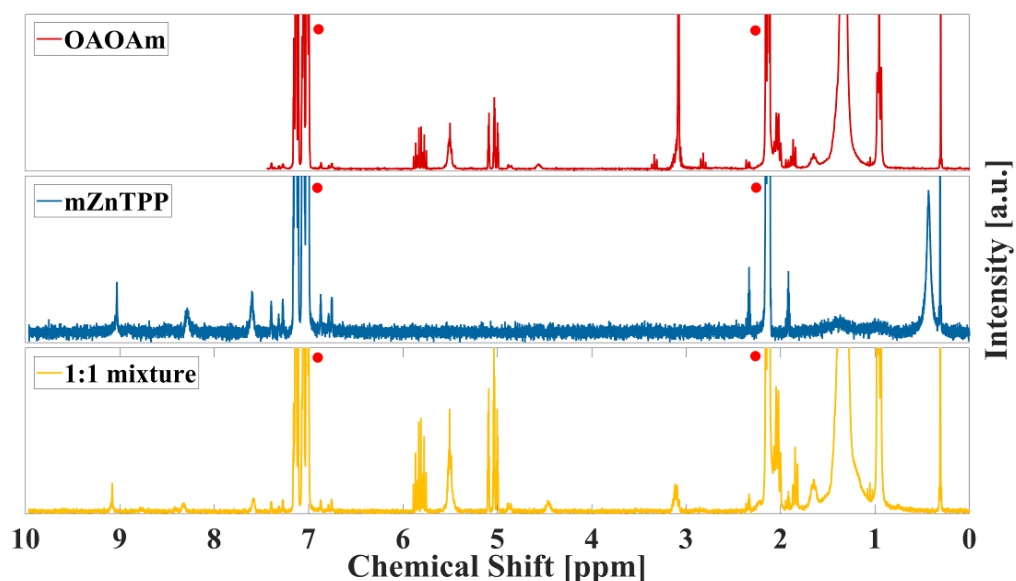

**Figure S4.** Overview NMR spectra of Top: Native CsPbBrI$_2$ functionalized with OA and OAm. Middle: Reference measurement of pure mZnTPP. Bottom: 1:1 stochiometric mixture of CsPbBrI$_2$ nanocrystals with mZnTPP. All spectra are taken in toluene, the solvent peaks are indicated by a red spot.

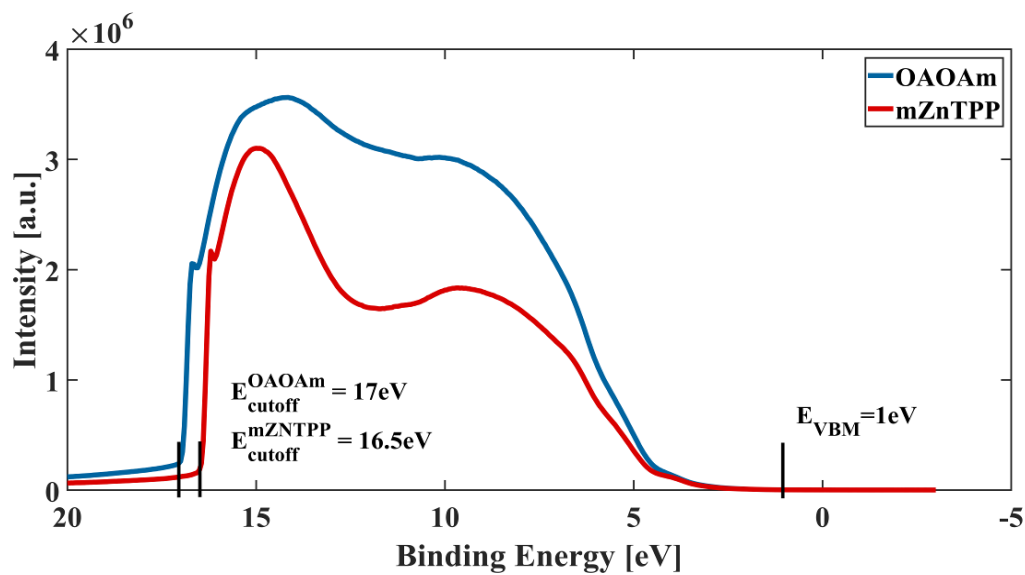

**Figure S5**. Overview UPS spectrum, indicated are the onset of the valence band maximum (VBM) as well as the cutoff energies.

**S5. Extinction coefficient of CsPbBrI$_2$**

The concentration of the perovskite solutions was determined by absorption spectroscopy, the extinction coefficient of CsPbBrI$_2$ thin films was taken as a reference.[1] Since CsPbBrI$_2$ nanocrystals with a size of 9nm are in the weak confinement regime (Exciton Bohr diameter ~10nm), the extinction coefficients derived from thin films are of sufficient accuracy. The absorbance can then easily be transformed to the molar extinction coefficient ε$_{mol}$ by multiplying with the molar volume.

$$\varepsilon_{mol}^{600} = \alpha_{600} * V_{mol} = 7263 \text{ cm}^{-1} \text{ M}^{-1}, \tag{S1}$$

with the absorption coefficient at 600nm α$_{600}$ and the molar volume of CsPbBrI$_2$ $V_{mol}$=0.14973 l/mol.

**S6. Photoluminescence spectra of mZnTPP and a stochiometric mixture with NCs**

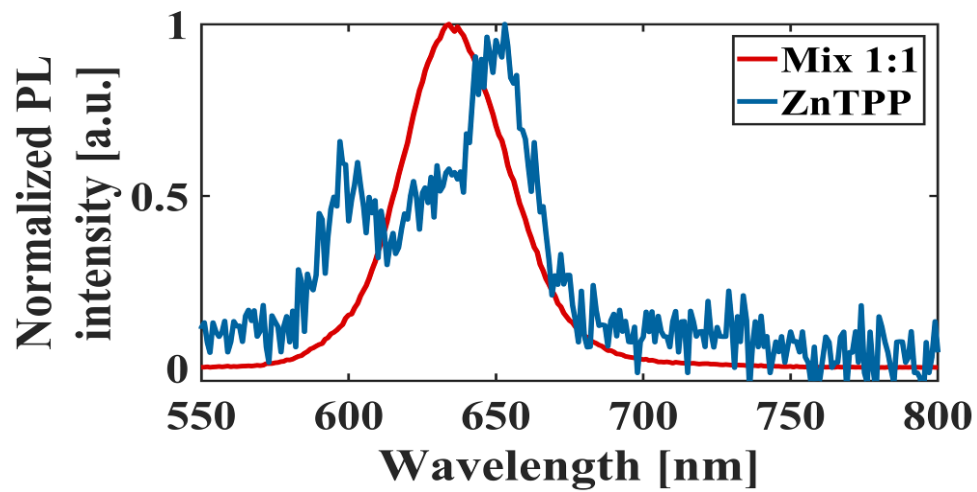

**Figure S5.** Normalized photoluminescence spectra of a 1:1 stochiometric mixture of $CsPbBrI_2$ NCs and mZnTPP (red) and pure mZnTPP (blue) in toluene.

## S7. Lifetime fitting model and parameters

As a general model for fitting the lifetime decay a triexponential function was used, as described by **Equation (S2)**, and the results and explicit fitting parameters are summarized in **Table S1**.

$$I(t) = I(0) + A_1 \exp\left\{-\frac{t-t_0}{\tau_1}\right\} + A_2 \exp\left\{-\frac{t-t_0}{\tau_2}\right\} + A_3 \exp\left\{-\frac{(t-t_0)}{\tau_3}\right\}, \quad \text{(S2)}$$

with Intensity $I$, preexponential factors $A_i$, time $t$ and lifetimes $\tau_i$.

**Table S1.** Complete fitting parameters for the triexponential function utilized to obtain the photoluminescence lifetimes. With Intensity $I$, preexponential factors $A_i$, lifetimes $\tau_i$ and the $R^2$ value to control the goodness of the fit.

|  |  | OAOAm | mZnTPP |
|---|---|---|---|
| Solution | $I(0)$ | 0.00590 | 0.00639 |
|  | $A_1$ | 0.08880 | 0.20708 |
|  | $\tau_1$ | 2.88499 | 2.24639 |
|  | $A_2$ | 0.49359 | 0.43820 |
|  | $\tau_2$ | 13.81641 | 12.93177 |
|  | $A_3$ | 0.34807 | 0.28029 |
|  | $\tau_3$ | 39.26681 | 34.06298 |
|  | $R^2$ | 0.99985 | 0.99983 |
|  |  | OAOAm | mZnTPP |
| Solid state | $I(0)$ | 0.00146 | 0.00152 |
|  | $A_1$ | 0.51995 | 0.58755 |
|  | $\tau_1$ | 1.09208 | 0.84141 |
|  | $A_2$ | 0.30650 | 0.30570 |
|  | $\tau_2$ | 6.58622 | 4.79796 |
|  | $A_3$ | 0.15721 | 0.15050 |
|  | $\tau_3$ | 22.46031 | 17.34084 |
|  | $R^2$ | 0.99971 | 0.99957 |

## S8. Temperature-dependent ionic conductivity and fitting results

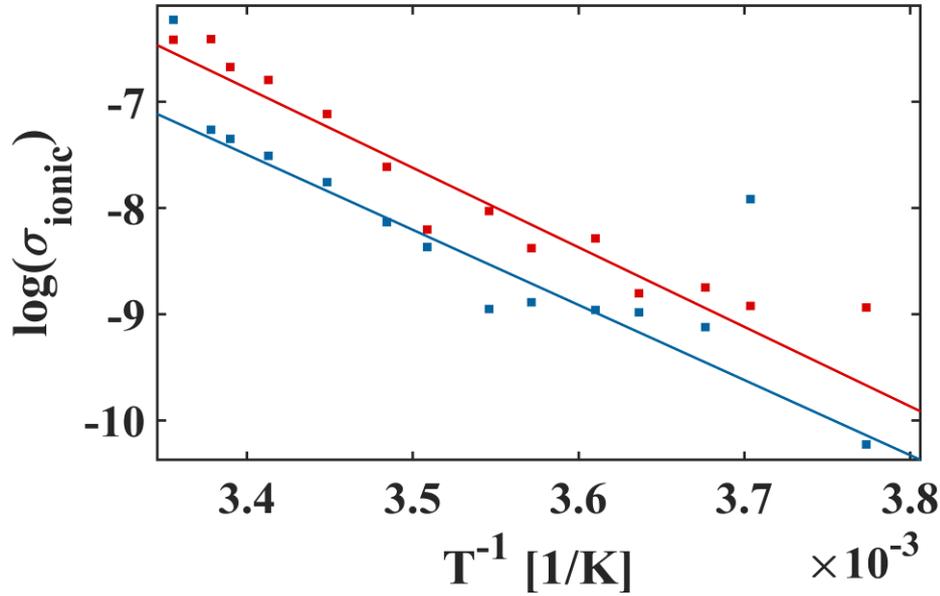

**Figure S6.** Temperature dependent ionic conductivity of native (red) and mZnTPP functionalized (blue) CsPbBrI$_2$ perovskites. The corresponding Arrhenius type NNH fits are shown as solid lines.

The results of the Arrhenius type NNH model are as follows:

Electronic contribution native

$$\sigma_{el}^{OAOAm}(T) = -17850 T^{-1} + 54.32 \ (higher\ T)$$
$$\sigma_{el}^{OAOAm}(T) = -7106 T^{-1} + 17.66 \ (lower\ T)$$

(S1)

The higher temperature part of the fit corresponds to the predominantly cubic and the lower temperature fit to the predominantly orthorhombic phase. [2,3]

Ionic contribution native $\quad\sigma_{ion}^{OAOAm}(T) = -6926 T^{-1} + 16.68$ (S2)

Electronic contribution mZnTPP $\quad\sigma_{el}^{mZnTPP}(T) = -11750\ T^{-1} + 33.28$ (S3)

Ionic contribution mZnTPP $\quad\sigma_{ion}^{mZnTPP}(T) = -7073\ T^{-1} + 16.55$ (S4)

## S9. Calculation of the charging energy

The conductivity of a nanocrystal thin film is dependent on the transfer integral β, which describes the dependence of the conductivity on the coupling between neighboring particles:

$$\beta \approx E_0 \times f(\Delta V, d_{NN}), \tag{S5}$$

with the activation energy $E_0$, containing the charging energy $E_C$ as well as fluctuations due to size distributions, and a contribution of the energy barrier with height $\Delta V$ and width $d_{NN}$.[4–6] To access the charging energy in Equation (7), the molecular polarizability and dielectric constant of different ligands was calculated with the Claussius-Mosotti equation in Equation (8).

$$\varepsilon_r = 1 + \frac{3 N_A \rho \alpha}{3 M_m \varepsilon_0 - N_A \rho \alpha}, \tag{S6}$$

with Avogadro's constant $N_A$, density $\rho$, molecular polarizability $\alpha$, molecular mass $M_m$ and vacuum permittivity $\varepsilon_0$.

On the other hand, the dielectric constant of the NCs is size-dependent and has to be adjusted following a generalized Penn model: [7]

$$\varepsilon_{Penn}(r) = 1 + \frac{\varepsilon_{bulk} - 1}{1 + \frac{18.05}{r^{1.8}}} \tag{S7}$$
$$\text{with } \varepsilon_{bulk} = 17$$

Finally, the charging energy was calculated following the modified Laikhtman-Wolf model in Equation (S10).[8]

$$E_C = \frac{e^2}{2(C_S + n C_m)}, \tag{S8}$$

with the self-capacitance

$$C_S^{-1} = \frac{1}{4\pi\varepsilon_0 r} \frac{\varepsilon_{QD} - \varepsilon_m}{\varepsilon_{QD}\varepsilon_m} + \frac{0.94}{4\pi\varepsilon_{QD}\varepsilon_0 r} \frac{\varepsilon_{QD} - \varepsilon_m}{\varepsilon_{QD} + \varepsilon_m} \tag{S9}$$

and the mutual capacitance

$$C_m = 2\pi\varepsilon_0 \frac{\varepsilon_{QD}\varepsilon_m}{\varepsilon_{QD} - \varepsilon_m} r \ln\left\{\frac{D}{D - 2r}\right\}. \tag{S10}$$

With the dielectric constant of the nanocrystal $\varepsilon_{QD}$, the dielectric constant of the ligand $\varepsilon_m$, the NC radius $r$ and the interparticle distance $D$.

Since the description of the charging energy, precisely the capacities, only hold true for spherical nanocrystals the calculated values had to be corrected by a constant factor which corresponds to the change to cubic morphologies. It has been shown for the unit cube that the capacitance must be adjusted by a factor of ~0.6607.[9]

Besides adjusting the capacities, a correction was made to properly apply the Penn model to the cubic NCs. The volume of the perovskite NCs was calculated and subsequently used to calculate the radius of a corresponding sphere with equivalent volume, which radius was inserted into the Penn model. Additionally, the value for the static dielectric constant was required, values for $CsPbI_3$ and $CsPbBr_3$ are known from literature.[10,11] Linear behavior of the mixed halide system was assumed and used to calculate the static dielectric constant.

**Table S2**: Static dielectric constants of iodide and bromide containing cubic perovskites used throughout this work.

|  | Static dielectric constant |
|---|---|
| $CsPbI_3$[11] | 18.1 |
| $CsPbBr_3$[10] | 15 |
| $CsPbBrI_2$ | 17 |

## S10. DFT calculations

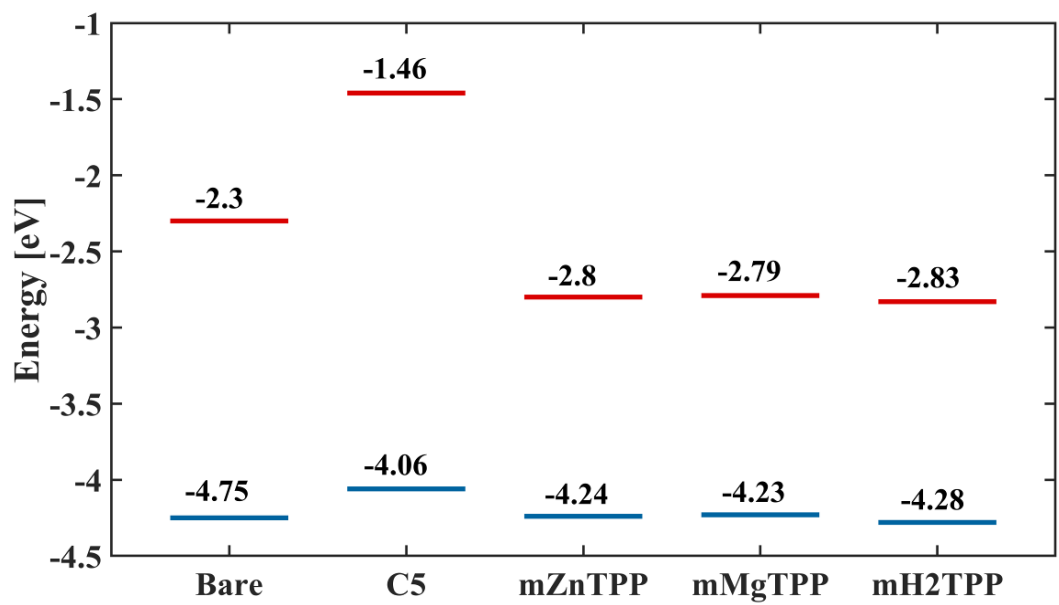

**Figure S7.** Energy levels of the frontier orbitals (highest occupied molecular orbital or HOMO (blue) and lowest unoccupied molecular orbital or LUMO (red)) obtained from calculations at the PBE/DZVP level for NCs functionalized with various ligands: aliphatic hexanoic acid (C5; mimic for oleic acid), and porphyrin derivatives: mZnTPP, mMgTPP and mH2TPP.

**Figure S8** demonstrates the frontier orbital energies of $CsPbBrI_2$ NCs having various surface ligands. While the experimental trend of a shift in the band gap energetics to lower energies upon exchanging the aliphatic ligand with semiconducting tetraphenylporphyrin derivatives as measured with UPS, is well reproduced computationally, the absolute values are quite different. This is expected, however, since semilocal DFT is well-known in literature to underestimate bandgaps. Additionally, calculations were run on nanocrystal models having 1.7 nm edge length due to computational limitations, whereas the particles used experimentally were roughly five times larger.

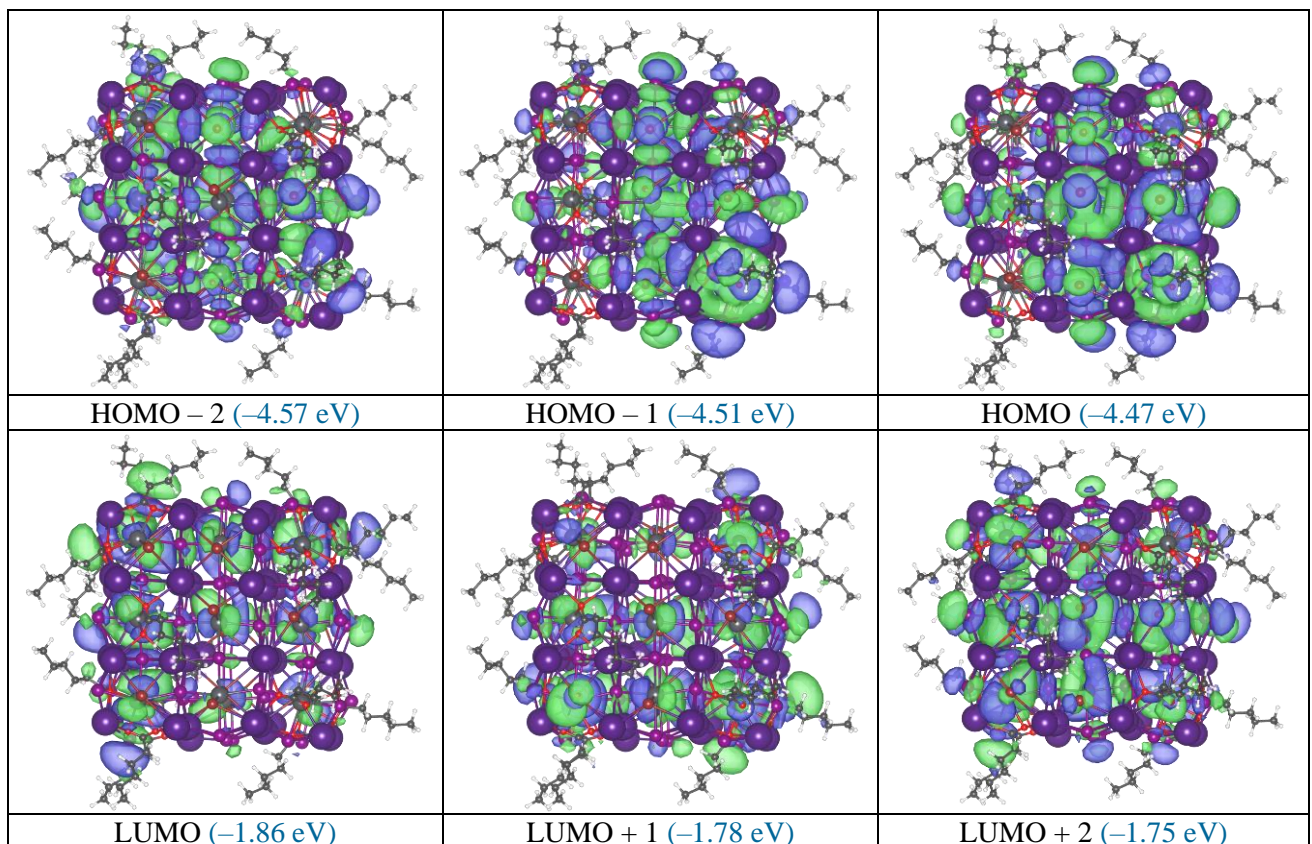

**Figure S8.** Calculated frontier molecular orbitals of hexanoic acid functionalized NCs.

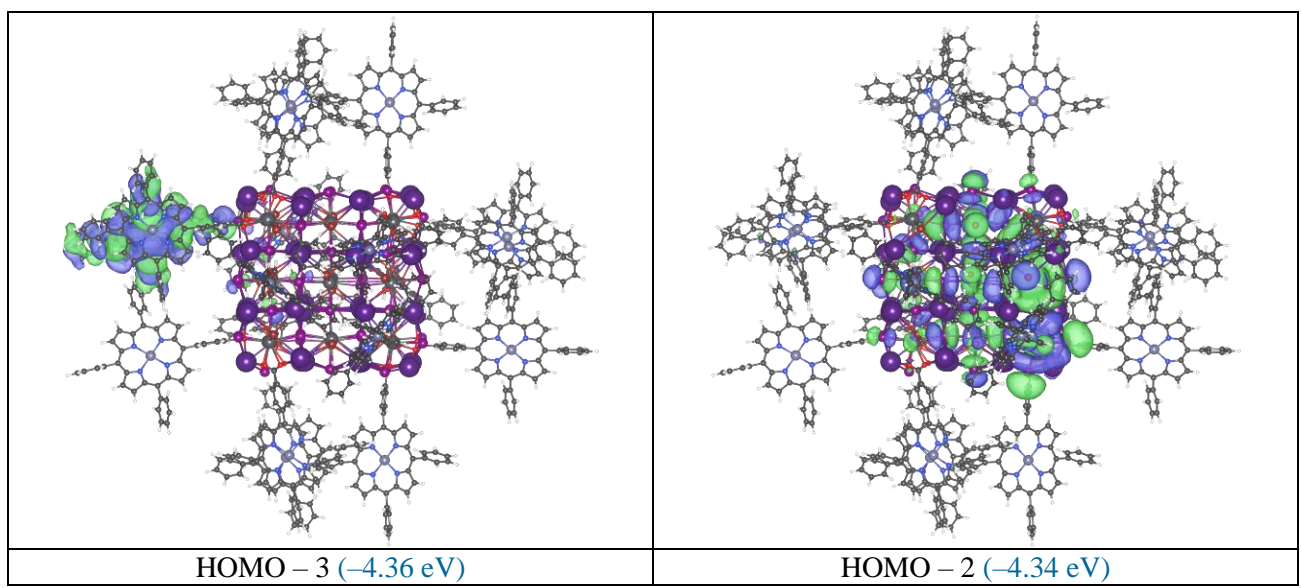

| HOMO – 3 (–4.36 eV) | HOMO – 2 (–4.34 eV) |

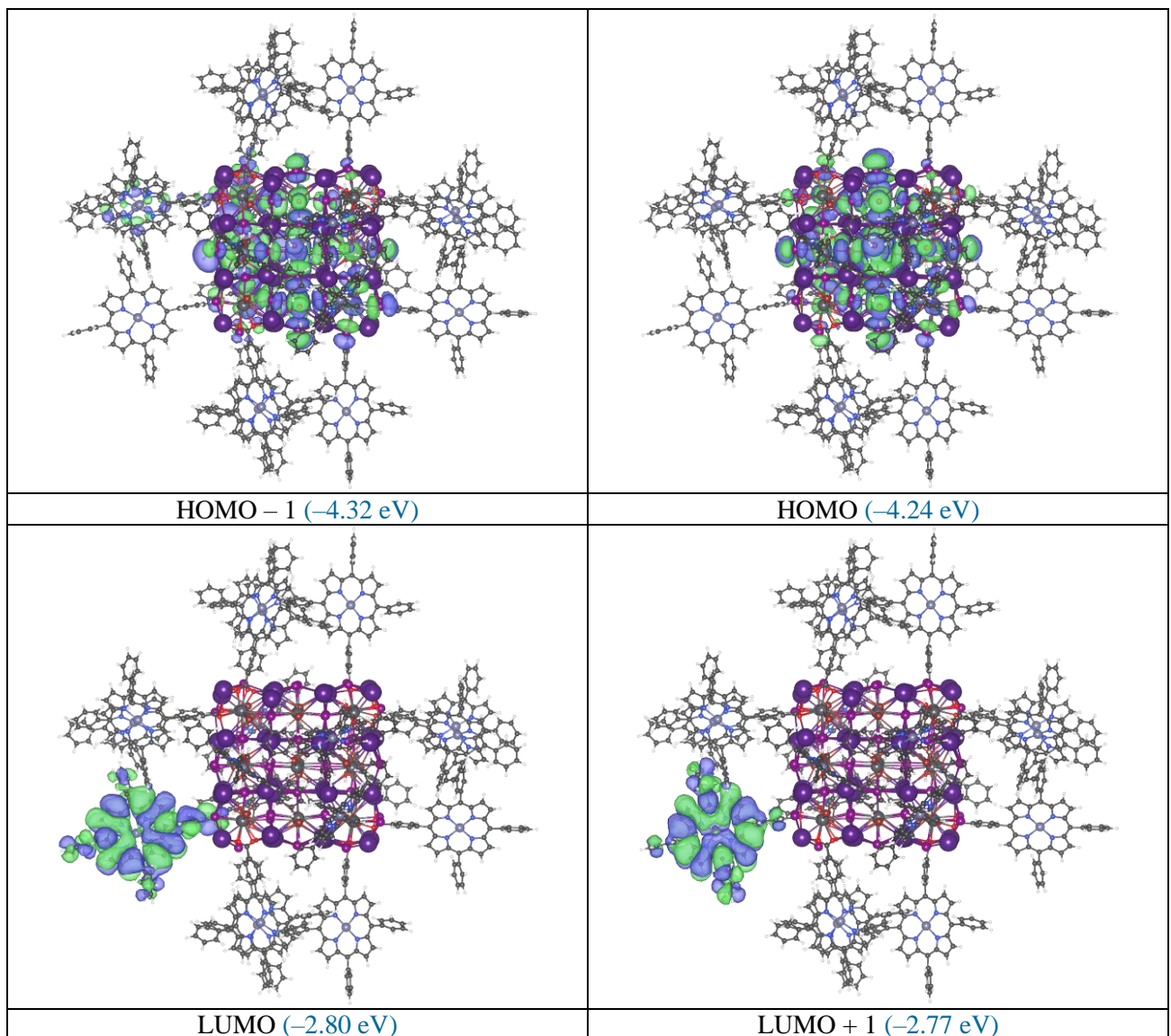

**Figure S9.** Calculated frontier molecular orbitals of mZnTPP functionalized NCs.

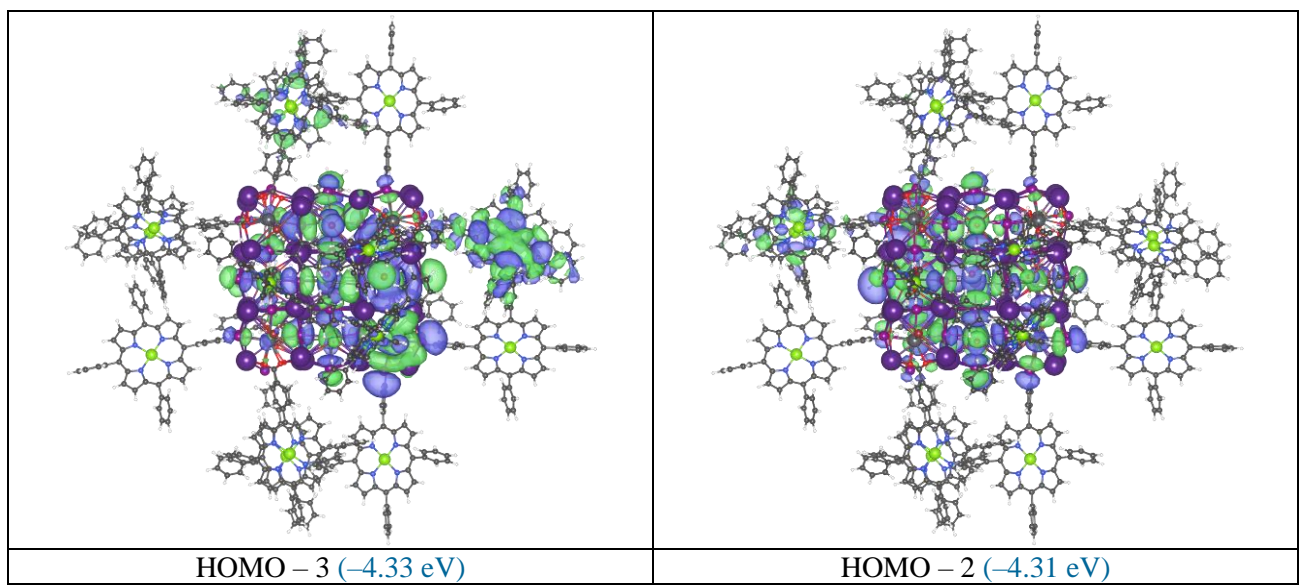

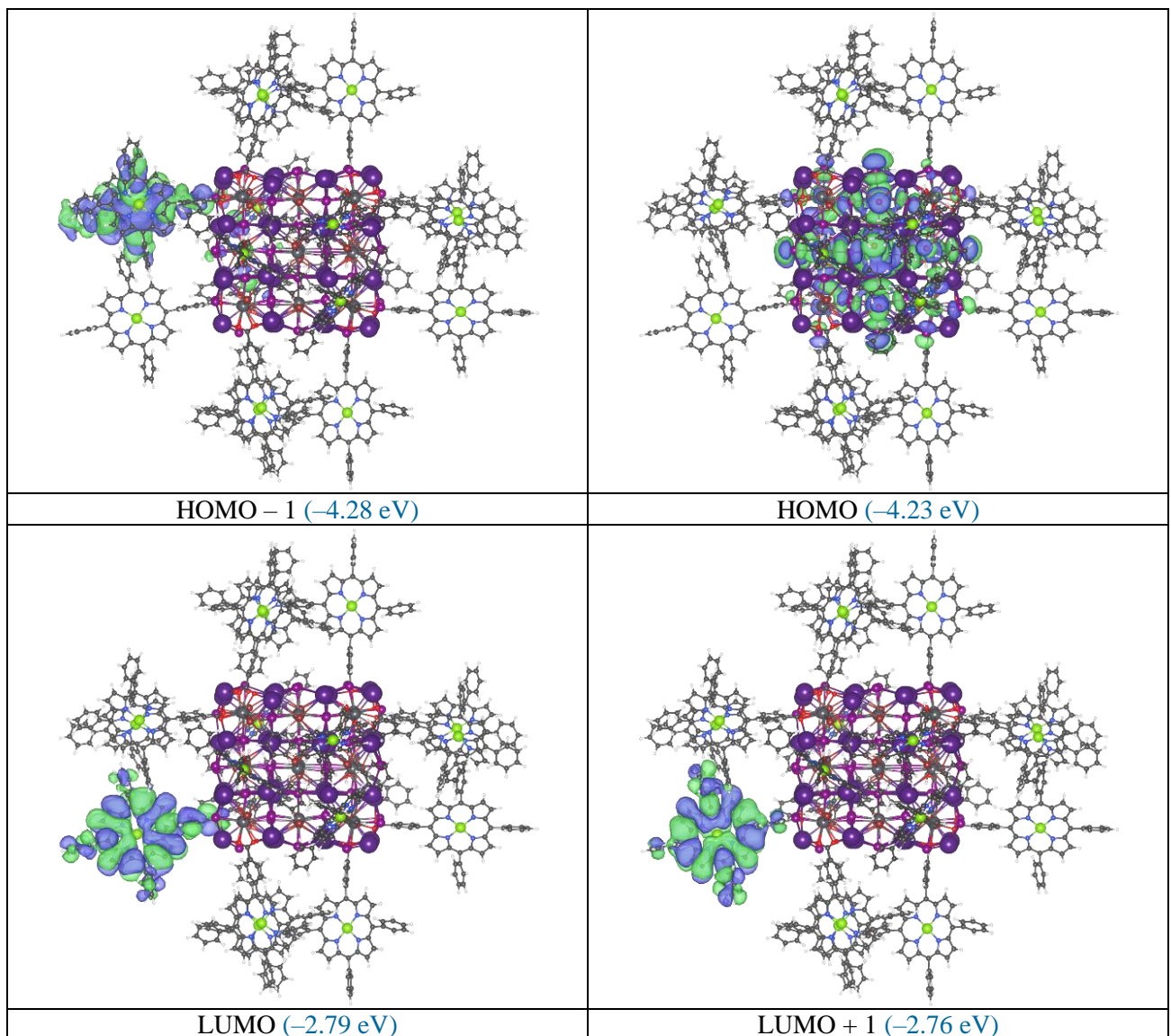

**Figure S10.** Calculated frontier molecular orbitals of mMgTPP functionalized NCs.

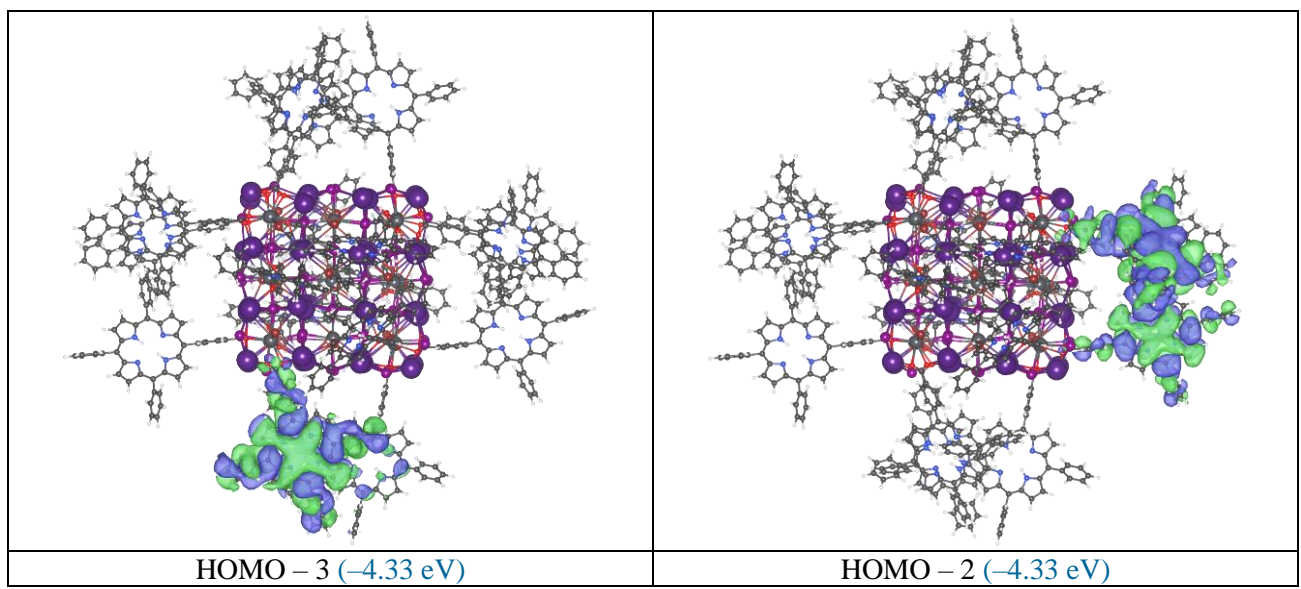

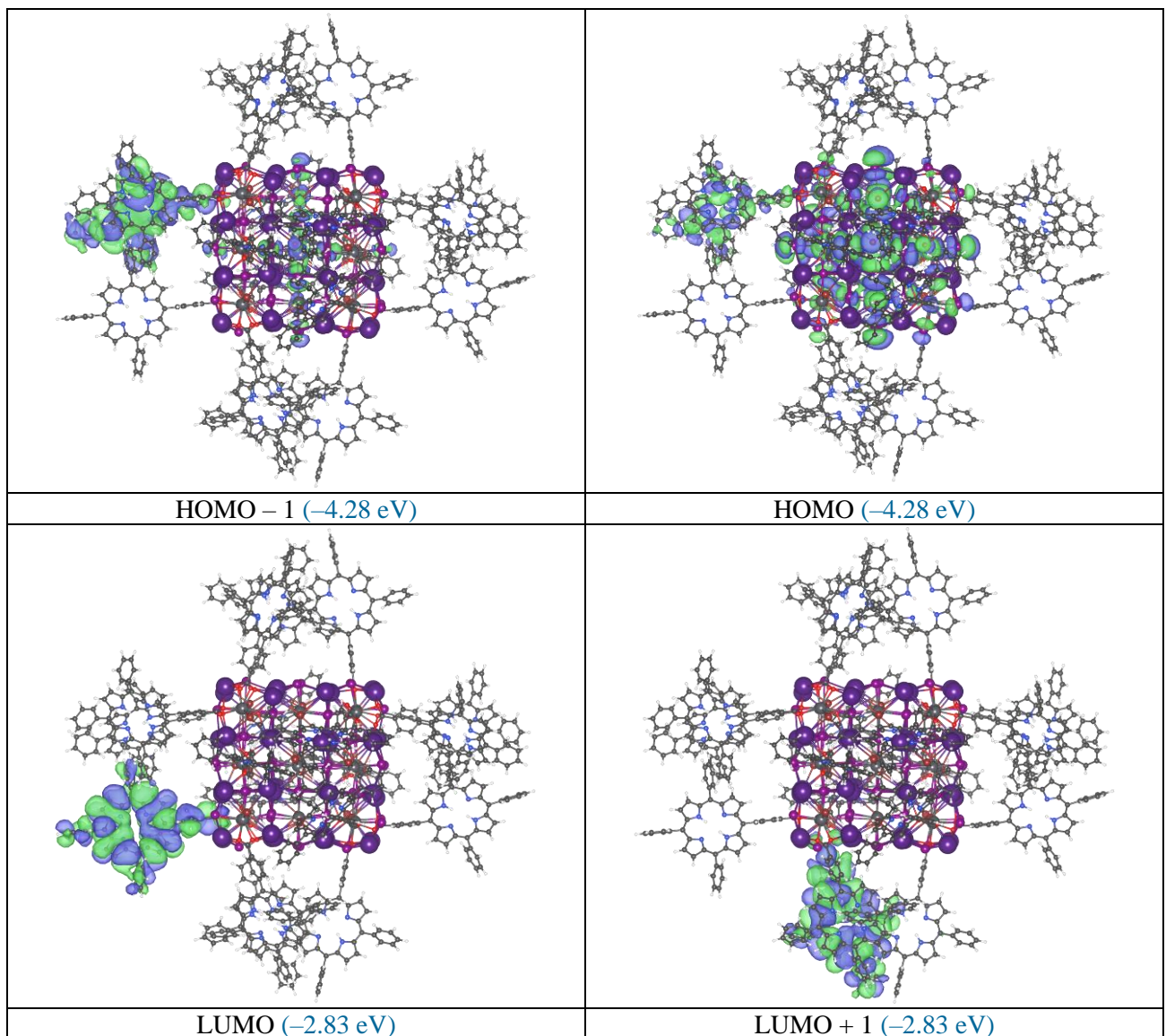

**Figure S11.** Calculated frontier molecular orbitals of mH2TPP functionalized NCs.

## S11. AFM of the film coverage on TFB

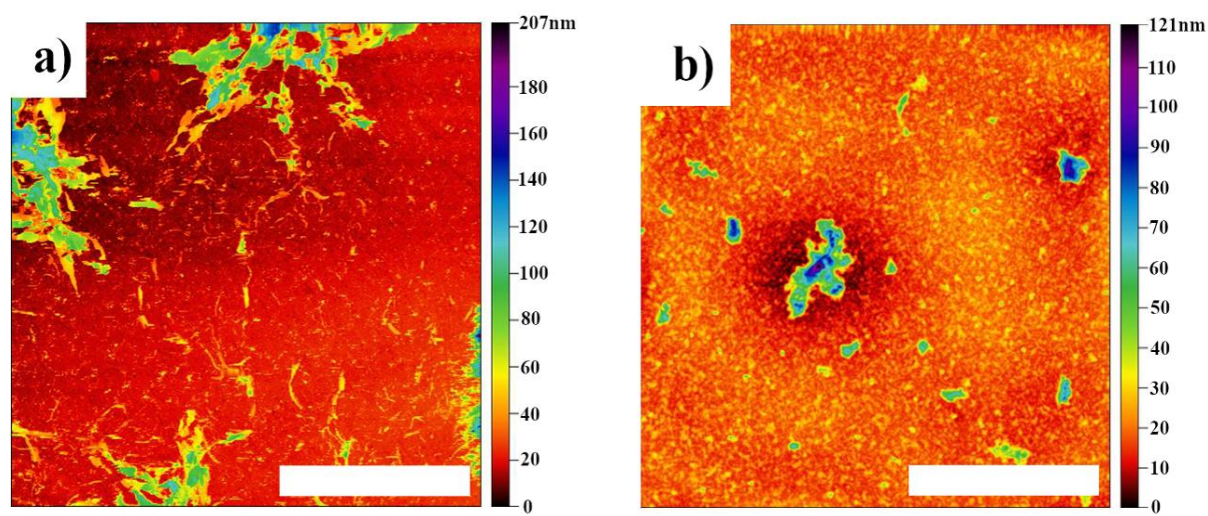

**Figure S12.** AFM images of the perovskite film coverage on a TFB coated substrate. a) Native CsPbBrI$_2$ nanoparticles on TFB. b) Exchanged CsPbBrI$_2$ nanoparticles on TFB. Scale bars are 4μm in both cases.